\documentclass[%
 reprint,
 amsmath,amssymb,
 aps,
]{revtex4-2}

\usepackage{graphicx}
\usepackage{dcolumn}
\usepackage{bm}
\usepackage{physics}
\usepackage{color}
\usepackage{float} 
\usepackage{orcidlink}

\usepackage{cancel}

\newcommand{\avrg}[1]{\left\langle #1 \right\rangle}
\newcommand{\bigo}{\mathcal{O}}

\makeatletter
\DeclareFontFamily{OMX}{MnSymbolE}{}
\DeclareSymbolFont{MnLargeSymbols}{OMX}{MnSymbolE}{m}{n}
\SetSymbolFont{MnLargeSymbols}{bold}{OMX}{MnSymbolE}{b}{n}
\DeclareFontShape{OMX}{MnSymbolE}{m}{n}{
    <-6>  MnSymbolE5
   <6-7>  MnSymbolE6
   <7-8>  MnSymbolE7
   <8-9>  MnSymbolE8
   <9-10> MnSymbolE9
  <10-12> MnSymbolE10
  <12->   MnSymbolE12
}{}
\DeclareFontShape{OMX}{MnSymbolE}{b}{n}{
    <-6>  MnSymbolE-Bold5
   <6-7>  MnSymbolE-Bold6
   <7-8>  MnSymbolE-Bold7
   <8-9>  MnSymbolE-Bold8
   <9-10> MnSymbolE-Bold9
  <10-12> MnSymbolE-Bold10
  <12->   MnSymbolE-Bold12
}{}

\let\llangle\@undefined
\let\rrangle\@undefined
\DeclareMathDelimiter{\llangle}{\mathopen}%
                     {MnLargeSymbols}{'164}{MnLargeSymbols}{'164}
\DeclareMathDelimiter{\rrangle}{\mathclose}%
                     {MnLargeSymbols}{'171}{MnLargeSymbols}{'171}
\makeatother
\begin{document}

\preprint{APS/123-QED}

\title{
Extended Yard Sale model of wealth distribution on Erd\H{o}s-R\'enyi random networks
}

\author{Nicolás Vazquez Von Bibow\,\orcidlink{https://orcid.org/0000-0003-0551-3561}}
\affiliation{Facultad de Matemática, Astronomía, Física y Computación (FAMAF), Universidad Nacional de Córdoba (UNC), Ciudad Universitaria 5000 Córdoba, Argentina, Grupo de Teoría de la Materia Condensada}
\affiliation{Instituto de Física Enrique Gaviola (IFEG-CONICET), Ciudad Universitaria, 5000 Córdoba, Argentina}
\author{Juan I. Perotti,\orcidlink{https://orcid.org/0000-0001-7424-9552}}%
\email{juan.perotti@unc.edu.ar}
\affiliation{Facultad de Matemática, Astronomía, Física y Computación (FAMAF), Universidad Nacional de Córdoba (UNC), Ciudad Universitaria 5000 Córdoba, Argentina, Grupo de Teoría de la Materia Condensada}
\affiliation{Instituto de Física Enrique Gaviola (IFEG-CONICET), Ciudad Universitaria, 5000 Córdoba, Argentina}




\date{\today}

\begin{abstract}
Excessive wealth concentration can undermine economic and social development. Random Asset Exchange (RAE) models provide valuable tools to investigate this phenomenon. 
Assuming that economic systems may operate optimally near the critical point of a continuous phase transition, the Extended Yard Sale (EYS) model introduced by Boghosian et al.~[Physica A 476, 15 (2017)] offers a compelling framework. 
This model captures the interplay between wealth redistribution and accumulation, exhibiting a continuous phase transition marked by a broad wealth distribution at criticality, separating a condensed phase---where a microscopic fraction of agents holds a macroscopic share of total wealth---from a distributed phase with a light-tailed wealth distribution.
While the original EYS model assumes fully connected interactions, this work introduces and studies a networked variant where agents interact over Erd\H{o}s-Rényi random networks. The analysis combines Monte Carlo simulations with Quenched Mean Field and Mean Field approximations, exploring a variety of interaction and taxation schemes. A scaling analysis shows that, although the networked model also undergoes a continuous phase transition, it leads to local wealth condensation rather than the global condensation found in the fully connected case.
These results deepen our understanding of wealth dynamics in structured populations and may help inform the development of more effective economic and social policies.
\end{abstract}

\maketitle


\section{
\label{intro}
Introduction
}

There is a concerning trend due to the increasing wealth concentration at a global scale, since it threats fair competition among economic agents, potentially undermining economic growth~\cite{piketty2014capital}. 
The question of whether there is an optimal level of wealth concentration is a subject of intense debate. 

Borrowing methods from statistical mechanics, numerous Random Asset Exchange (RAE) models have been developed to study wealth distributions~\cite{angle1986surplus, ispolatov1998wealth, dragulescu2000statistical, yakovenko2009colloquium, chakraborti2011econophysicsI, chakraborti2011econophysicsII, chakrabarti2013econophysics, greenberg2024twentyfive}.
Notably, several of these models predict the emergence of a wealth-condensed state, in which a microscopic fraction of agents---referred to as oligarchs---accumulate a macroscopic share of the total wealth~\cite{bouchad2000wealth, boghosian2017oligarchy, boghosian2015theorem, cardoso2023why}.
Wealth condensation is widely regarded as an undesirable outcome. 
This becomes particularly evident in the theoretical limit where all wealth is concentrated in the hands of a single agent, effectively halting economic activity.
Even at intermediate levels of condensation, the situation remains problematic, since wealthy agents may leverage their accumulated capital not necessarily through superior economic performance, but rather by securing privileged positions and opportunities within society.

The Yard Sale (YS) model~\cite{chakraborti2002distribution,hayes2002computing} and its variants are particular cases of RAE models, in which the exchanged wealth during interactions is a fraction of the poorer's agent.
YS models are among of the most widely used frameworks to study wealth distributions and wealth condensation.
Various extensions of the YS model have incorporated mechanisms of wealth redistribution~\cite{boghosian2014fokker,
boghosian2014kinetics}, inflation, production, taxation~\cite{boghosian2014kinetics,nener2022study}, debt~\cite{li2019affine}, saving propensities~\cite{chatterjee2004pareto}, risk aversion~\cite{iglesias2004correlation,nener2021optimal}, biases~\cite{moukarzel2007wealth,
boghosian2017oligarchy} or rational behavior~\cite{nener2021wealth}.
In particular, the recently introduced Extended YS (EYS) model by Boghosian et al.~\cite{boghosian2017oligarchy,li2019affine} combines two biases, the mechanisms of wealth redistribution that favors the poor and Wealth Attained Advantage (WAA) that favors the rich during interactions.
Notably, the EYS model exhibits a critical redistribution rate at which a continuous phase transition emerges between a wealth-condensed phase and a disordered phase.  
This is noteworthy because such a feature is characteristic of a wide range of complex systems~\cite{bak1996nature, perotti2009emergent, chialvo2010emergent, mora2011biological,  
tang2017critical, cavagna2018physics, munoz2018colloquium, li2019affine, chialvo2020controlling,  
zamponi2022universal}.

Euler introduced the concept of graphs studying the problem of the seven bridges of K\"onigsberg in 1736.
Most early works in physics considered regular graphs such as lattices or the Cayley tree.
In the 50's Paul Erd\H{o}s and Albert R\'enyi developed the theory of random graphs, which imitate the disorder observed in real networks.
Still, real networks are not regular, nor purely random, but are complex.
This fact was noticed by Watts and Strogatz~\cite{watts1998collective} and Barab\'asi and Albert~\cite{barabasi1999emergence} who respectively introduced the concepts of small-world and scale-free networks, giving birth to the modern theory of complex networks~\cite{newman2018networks}.

In real-world settings, social agents typically interact repeatedly with a relatively small set of neighbors rather than with the entire population~\cite{onnela2007structure}. Nevertheless, most variants of the YS model assume fully connected interactions, where any pair of agents can trade, which is an unrealistic assumption.
Although some versions of the YS model have been adapted to networked settings~\cite{vazquez2010wealth,bustos2012yard,lee2023scaling}, these remain comparatively less explored. Studying networked variants is crucial, as interaction structure can profoundly affect wealth dynamics.
For example, a recent preprint by B\"orgers and Greengard demonstrates that global wealth condensation does not occur in a broad class of networked variants of YS models, but it is replaced by a notion of local wealth condensation where wealthy agents cannot be neighbors~\cite{borgers2024local}.

In this work, a networked variant of the EYS model is introduced and analyzed through Monte Carlo (MC) simulations, Quenched Mean Field (QMF) approximations, and a Mean Field (MF) analytical approach. 
In Sec.~\ref{theory}, the original EYS model is revisited, and key concepts from network theory are reviewed.
The networked EYS model is then formulated, incorporating different taxation and interaction schemes, and the corresponding QMF approximation is developed.
In Sec.~\ref{results}, numerical results obtained from MC simulations and QMF approximations are presented for both fully connected and Erd\H{o}s–R\'enyi random networks.
A scaling analysis is carried out to characterize the continuous phase transition, and the MF theory is subsequently introduced and examined.
Finally, the main conclusions are provided in Sec.~\ref{conclusion}.

\section{
\label{theory}
Theory
}

\subsection{The Extended Yard Sale model}

In the Extended Yard Sale model (EYS)~\cite{boghosian2017oligarchy}, $N$ agents exchange wealth following a stochastic process where $w_i(t)$ denotes the wealth of agent $i$ at time $t$.
The total wealth is conserved, so $W=\sum_i w_i(t)$ for all $t$.
At each time step $t\to t+\Delta t$, two agents $i$ and $j$ are chosen at random to exchange wealth and pay or receive benefits from the tax system according to the rule
\begin{eqnarray}
\label{eq:1} 
w_i(t+\Delta t) 
&=&
w_i(t) 
+ 
\chi
\Delta t\,\bigg(\frac{W}{N}-w_i(t)\bigg) 
\\
&&
+ 
\sqrt{\gamma \Delta t}\,
\eta_{ij}(t)\,
\min\{w_i(t),w_j(t)\} 
\nonumber
\end{eqnarray}
for agent $i$ and an analogous rule for agent $j$.
The second term of Eq.~\ref{eq:1} represents the social benefit earned minus the tax payed by agent $i$.
Its action redistributes wealth from the rich ($w_i>W/N$) to the poor ($w_i<W/N$) agents.
The third term represents the amount of wealth exchanged between $i$ and $j$ in the interaction.
By definition of the YS model, it is a fraction of the wealth of the poorer agent, so
\begin{equation}
\label{eq:2}
-1\leq \sqrt{\gamma \Delta t}\, \eta_{ij}(t)\leq 1
\end{equation}
must be satisfied.
The antisymmetric matrix $\eta(t)$ is a stochastic variate of entries $\eta_{ij}(t)=-\eta_{ji}(t)\in \{-1,1\}$ representing the direction of wealth flow.
Each upper-diagonal entry independently takes a value from a distribution with mean
\begin{equation}
\label{eq:3}
\avrg{\eta_{ij}} = \zeta N
\sqrt{\frac{\Delta t}{\gamma}}\, \frac{w_i-w_j}{W}
\end{equation}
and second moment
$\avrg{\eta_{ij}^2}=1$.
The bias implied by Eq.~\ref{eq:3} is called the Wealth Attained Advantage (WAA) and always favors the richest agent of the transaction.

In the original YS model, which includes neither redistribution nor WAA mechanisms, the system invariably evolves toward a winner-takes-all absorbing state of wealth condensation, in which a single agent eventually acquires all the wealth as $t \to \infty$. To mitigate this extreme outcome, Boghosian et al. introduced the redistribution mechanism, which completely eliminates the condensed phase~\cite{boghosian2014kinetics}. To restore condensation, they subsequently incorporated the WAA mechanism alongside redistribution, thereby counteracting the equalizing effects of the latter~\cite{boghosian2017oligarchy}. The inclusion of both mechanisms defines the EYS model, which displays a stationary condensed phase for $\chi < \zeta$ and a distributed phase for $\chi > \zeta$, separated by a continuous transition at $\chi = \zeta$ where the long-time wealth distribution, $P(w) = \frac{1}{N} \sum_i \delta(w - w_i)$, becomes heavy-tailed.

Equation~\ref{eq:3} is consistent with a distribution of probabilities
$$
P(\eta_{ij})
=
\frac{1}{2}
\bigg(
1
+
\eta_{ij}
\avrg{\eta_{ij}}
\bigg)
$$
given that $-1\leq\avrg{\eta_{ij}}\leq 1$.
This is consistent only if
\begin{equation}
\label{eq:4}
\zeta N \sqrt{\frac{\Delta t}{\gamma}} \leq 1.
\end{equation}
The conditions of Eqs.~\ref{eq:2}~and~\ref{eq:4} are simultaneously satisfied if, for example, 
$\Delta t=\gamma N^{-2}$ and $\zeta=1$.
The factor $N$ in Eq.~\ref{eq:3} is necessary in order to obtain finite critical points in the thermodynamic limit $N\to \infty$.

\subsection{
\label{sec:networks}
Networks
}

The structure of networks is usually represented by graphs, which in turn are composed by a set of nodes and a set of links.
The number of directed links of source $j$ and target $i$ of a network of $N$ nodes in $\{1,...,N\}$ can be represented by the entry $a_{ij}\in \{0,1,2,...\}$ of a corresponding adjacency matrix $a$~\cite{newman2018networks}.
The number of outgoing and incoming links of node $i$ are denoted by $k_i=\sum_j a_{ij}$ and $q_i=\sum_j a_{ji}$ respectively, $M=\sum_{ij} a_{ij}=\sum_i k_i = \sum_j q_j=N\bar{k}=N\bar{q}$ denotes the number of directed links, and $\bar{k}$ and $\bar{q}$ the average out- and in-degree of the nodes of the network~\footnote{Note that, usually, $2M=\sum_{ij} a_{ij}$ is used for nondirected networks.
Here $M=\sum_{ij} a_{ij}$ is preferred since it works for both, directed and non-directed networks.}.
Non-directed networks are represented with symmetric adjacency matrices, in which case the number of non-directed links is $M/2$, $k_i=q_i$ and, by convention, $a_{ii}\in \{0,2,4,...\}$ for each $i$. 

Many network models can be represented by a distribution of probabilities $P(a|\theta)$ for adjacency matrices $a$ where $\theta$ is a vector of model parameters~\cite{newman2018networks}.
For instance, in the case of Erd\H{o}s-R\'enyi networks, $P(a|N,p)=\prod_{i<j} p^{a_{ij}}(1-p)^{1-a_{ij}}[a_{ij}=a_{ji}]$ where $p=MN^{-1}(N-1)^{-1}=\bar{k}/(N-1)\in [0,1]$ is the expected density of links of the generated networks and $[...]$ denotes Iverson's bracket.
In some other cases, like in the Barab\'asi-Albert network model, an explicit formula for the distribution $P(a|\theta)$ cannot be easily obtained, if possible, but it still exists.

\subsection{
\label{sec:EYSM_on_nets}
The EYS model on networks
}

To obtain a generalization of the EYS model on networks, consider a given adjacency matrix $a$, and redefine the change in wealth $\Delta w_i(t) = w_i(t+\Delta t)-w_i(t)$ that occurs in a time step $t\to t+\Delta t$ with the expression
\begin{eqnarray}
\label{eq:5}
\Delta w_i(t)
&=&
\chi
\Delta t\,
\bigg(
-
b_i(t)
w_i(t)
+
\frac{1}{N}
\sum_j
b_j(t)
w_j(t)
\bigg) 
\\
&&
+ 
\sum_j
c_{ij}(t)
\sqrt{\gamma \Delta t}\,
\eta_{ij}(t)\,
\min\{w_i(t),w_j(t)\} 
.
\nonumber
\end{eqnarray}
Here, the vector $b(t)$ of entries $b_i(t)\in \{0,1\}$ is a stochastic variable used to represent if a taxation event occurs to agent $i$ in the time interval $[t,t+\Delta t)$.
Similarly, the symmetric matrix $c(t)$ of entries $c_{ij}(t)\in \{0,1\}$ is a stochastic variable used to indicate if an interaction occurs between agents $i$ and $j$ within the same time interval.
The term $N^{-1}\sum_j b_jw_j$ is introduced to replace $W/N$ in Eq.~\ref{eq:1} to enforce the conservation condition $\sum_i \Delta w_i(t)=0$ of the networked variant of the model. 

\subsection{
\label{sec:QMF_approx}
QMF approximation for the EYS model on networks
}

\begin{widetext}
\begin{eqnarray}
\label{eq:6}
\avrg{\Delta w_i}_{sw|a}
&=&
\sum_s
P(s|a)
\bigg\{
\sum_{b}
P(b|s,a)\,
\chi
\Delta t
\int dw_i\,
P(w_i|a)
\bigg(
-
b_iw_i
+
\frac{1}{N}
\sum_j
b_j
w_i
\bigg)
\\
&&
+
\sum_{c}
P(c|s,a)
\sum_j
c_{ij}
\sqrt{\gamma \Delta t}
\int
d\eta_{ij}\,
dw_{i}\,
dw_{j}\,
P(\eta_{ij}|w_i,w_j)
P(w_i,w_j|a)
\,
\eta_{ij}
\min\{w_i,w_j\}
\bigg\}
\nonumber
\\
&=&
\chi
\Delta t
\,
\bigg(
-
\avrg{w_i}_{w|a}
\avrg{b_i}_{s|a}
+
\frac{1}{N}
\sum_j
\avrg{w_j}_{w|a}
\avrg{b_j}_{s|a}
\bigg)
\nonumber
\\
&&
+
\zeta N
\Delta t
\sum_j
\avrg{c_{ij}}_{s|a}
\int
dw_{i}\,
dw_{j}\,
P(w_i,w_j|a)
\,
\bigg(
\frac{w_i-w_j}{W}
\bigg)
\min\{w_i,w_j\}
\bigg\}
\nonumber
\end{eqnarray}
\end{widetext}

In what follows, a so-called Quenched Mean Field (QMF) approximation for the dynamics of the networked variant of the EYS model of Eq.~\ref{eq:5} is introduced~\cite{pastor2015epidemic}.
The expected change of $\Delta w_i$ is obtained by integrating Eq.~\ref{eq:5} with respect to the stochastic variables $b$, $c$, $w$ and $\eta$.
The result is shown in Eq.~\ref{eq:6}, where the explicit dependency of the stochastic variables $w$, $\eta$, $s$, $b$ and $c$ with time $t$ is omitted for simplicity. 
Notation $\avrg{\Delta w_i}_{sw|a}$ represents the expected value of $\Delta w_i$ over $s$ and $w$ while keeping $a$ fixed.
A new stochastic variable $s$ is introduced to account for the common conditional cause behind the joint realization of the interaction and taxation events occurring within the time interval $[t,t+\Delta t)$.
The expected value 
$$
\beta_i
:=
\avrg{b_i}_{s|a}
=
\sum_s P(s|a)\sum_b P(b|s,a) b_i
$$ 
is the probability for agent $i$ to be taxed per time step and the expected value 
$$
\kappa_{ij}
:=
\avrg{c_{ij}}_{s|a}
=
\sum_s P(s|a)\sum_c P(c|s,a) c_{ij}
$$
is the probability for agents $i$ and $j$ to interact per time step.
The Poissonian limit is assumed.
Therefore, multiple taxation and interaction events within a time step will contribute negligibly as $\Delta t$ goes to zero.

To obtain the QMF approximation of the networked variant of the EYS model, assume that
$$
P(w_i,w_j|a)
\approx
P(w_i|a)P(w_j|a)
$$
and that each $P(w_i|a)$ is highly peaked around the mean $\avrg{w_i}_{w|a}$.
In this way, Eq.~\ref{eq:6} transforms into 
\begin{eqnarray}
\label{eq:7}
\frac{d\avrg{w_i}_{w|a}}{dt}
&\approx &
\chi
\bigg(
-
\avrg{b_i}_{s|a}
\avrg{w_i}_{w|a}
+
\frac{1}{N}
\sum_j
\avrg{b_j}_{s|a}
\avrg{w_j}_{w|a}
\bigg)
\nonumber
\\
&&
\;\;\;\;
+
\zeta
N
\sum_j
\avrg{c_{ij}}_{s|a}
\bigg(
\frac{\avrg{w_i}_{w|a}-\avrg{w_j}_{w|a}}{W}
\bigg)
\nonumber
\\
&&
\;\;\;\;
\;\;\;\;
\times
\min\{\avrg{w_i}_{w|a},\avrg{w_j}_{w|a}\}
\end{eqnarray}
where it is also assumed that
$$
\frac{\avrg{\Delta w_i}_{sw|a}}{\Delta t}
\to
\frac{d\avrg{w_i}_{w|a}}{dt}
$$
as $\Delta t\to 0$ (i.e. as $N\to \infty$).
For convenience, and without loss of generality, the definitions $T := \chi / \zeta$, $x_i := \langle w_i \rangle_{w|a} / W$, and a rescaled time variable $\tau := \zeta N t$ are introduced.
After neglecting corrections due to correlations, the ordinary differential equation (ODE) in Eq.~\ref{eq:7} is simplified to:
\begin{eqnarray}
\label{eq:8}
\dot{x}_i
&=&
\frac{T}{N}
\bigg(
-\beta_i x_i+\frac{1}{N}\sum_j \beta_j x_j
\bigg)
\\
&&
+
\sum_j
\kappa_{ij}
(x_i-x_j)
\min\{x_i,x_j\}
\notag
\end{eqnarray}
where $\dot{x}_i = dx_i/d\tau$.
It can be shown that the conservation law $\sum_i \dot{x}_i=0$ holds if $\kappa_{ij}=\kappa_{ji}$ for all $i,j$.
As later shown, $T$ plays a role analogous to that of a temperature, in the sense that a disordered phase emerges at high temperatures and an ordered or condensed phase emerges at low temperatures.
The reader should keep in mind, however, that the systems being modeled are out of thermodynamic equilibrium and, as a consequence, initial conditions play an important role, as shown in Appendix~\ref{appA}.
Equation~\ref{eq:8} summarizes the QMF approximation of the EYS model on networks.

\vspace{.5cm}

\subsection{
\label{sec:IM_TM}
Interaction and taxation modes over networks
}

Different taxation and interaction modes can be considered.
Each interaction-taxation event is represented by a tuple $s=(uv,l,m)$ where the pair $uv$ denotes a directed link chosen from a network $a$ representing an interaction initiated by an agent $u$ over an agent $v$, and $l$ and $m$ represent the pair of agents that are being taxed.
Without loosing generality, at each event two agents are taxed simultaneously to unify different taxation modes within the same representation.
The outcomes of $uv$, $l$ and $m$ affect the statistics of $b$ and $c$. 
Specifically, $P(c_{ij}=1|s,a) =P(c_{ij}=1|uv)= \delta_{ij,uv}+\delta_{ij,vu}$, $P(c_{ij}=0|s,a)=1-P(c_{ij}=1|s,a)$, $P(b_i=1|s,a)=P(b_i=1|l,m)=\delta_{il}+\delta_{im}$ and $P(b_i=0|s,a)=1-P(b_i=1|s,a)$.

Two interaction modes are considered.
In Interaction Mode A (IMA), a directed link $uv$ is chosen uniformly at random among the $M$ available.
This occurs with probability 
$$
P(uv|a)
=
\frac{a_{uv}}{M}
$$
and, in consequence
\begin{eqnarray}
\label{eq:9}
\avrg{c_{ij}}_{s|a}
&=& 
\sum_{uv} P(uv|a)\sum_{c_{ij}\in\{0,1\}}P(c_{ij}|uv)
\,c_{ij}
\\
&=& 
\frac{a_{ij}+a_{ji}}{M}
.
\nonumber
\end{eqnarray}
In Interaction Mode B (IMB), an agent $u$ is chosen uniformly at random and then another agent $v$ is uniformly chosen at random among the $k_u$ neighbors of $u$.
In this way, 
$$P(uv|a)=P(v|ua)P(u|a)=\frac{a_{uv}}{k_u}\frac{1}{N},$$
so, by Eq.~\ref{eq:9}
\begin{eqnarray}
\avrg{c_{ij}}_{s|a}
&=& 
\sum_{uv} P(uv|a)\sum_{c_{ij}\in\{0,1\}}P(c_{ij}|uv)c_{ij}
\nonumber
\\
&=& 
\sum_{uv} \frac{1}{N}\frac{a_{uv}}{k_u}
\big(
\delta_{uv,ij}+\delta_{vu,ij}
\big)
\nonumber
\\
&=& 
\frac{1}{N}
\bigg(
\frac{a_{ij}}{k_i}
+
\frac{a_{ji}}{k_j}
\bigg)
.
\nonumber
\end{eqnarray}

For each interaction mode, two taxation modes are also considered.
In Taxation Mode A (TMA), the agents $u$ and $v$ participating in the interaction $uv$ are the ones being taxed.
Formally, $l=u$ and $m=v$, so $P(l,m|uv)=P(l|uv)P(m|uv)=\delta_{lu}\delta_{mv}$.
Therefore
\begin{eqnarray}
\label{eq:10}
\avrg{b_i}_{s|a}
&=&
\sum_{uv}
P(uv|a)
\sum_{lm}
P(l,m|uv)
\sum_{b_i}
P(b_i|l,m)
b_i
\nonumber
\\
&=&
\sum_{uv}
P(uv|a)
\sum_{lm}
\delta_{lu}
\delta_{mv}
\big(
\delta_{il}
+
\delta_{im}
\big)
\nonumber
\\
&=&
\sum_{uv}
P(uv|a)
\big(
\delta_{iu}
+
\delta_{iv}
\big)
.
\end{eqnarray}
Hence, under IMA, Eq.~\ref{eq:10} reduces to
\begin{eqnarray}
\avrg{b_i}_{s|a}
&=&
\sum_{uv}
\frac{a_{uv}}{M}
\big(
\delta_{iu}
+
\delta_{iv}
\big)
\nonumber
\\
&=&
\frac{k_i+q_i}{M}
.
\nonumber
\end{eqnarray}
On the other hand, under IMB, Eq.~\ref{eq:10} reduces to
\begin{eqnarray}
\avrg{b_i}_{s|a}
&=&
\sum_{uv}
\frac{1}{N}
\frac{a_{uv}}{k_u}
\big(
\delta_{iu}
+
\delta_{iv}
\big)
\nonumber
\\
&=&
\frac{1}{N}
\big(
1
+
h_{i}
\big)
\approx 
\frac{1}{N}
+
\frac{q_i}{M}
\nonumber
\end{eqnarray}
where $h_i = \sum_u a_{ui}/k_u \approx \sum_u a_{ui}/\bar{k} = q_i/\bar{k}$ if $1/k_u\approx 1/\bar{k}$ is assumed for all $u$ in the neighborhood of $i$.
In Taxation Mode B, the pair of agents being taxed are uniformly chosen at random.
Hence, $P(l,m|uv)=P(l|uv)P(m|uv)=N^{-2}$ and, therefore
\begin{equation}
\avrg{b_i}_{s|a}
=
\sum_{uv}P(uv|a)\sum_{lm}\frac{1}{N^2}(\delta_{il}+\delta_{im})=\frac{2}{N}
\nonumber
\end{equation}
for both interaction modes.

Although other reasonable interaction and taxation modes can be considered, this work focuses on IMA/IMB and TMA/TMB for simplicity.
The study of alternatives is left open for future work.
Note that TMA is consistent with the taxation of transactions and TMB with an homogeneous taxation rate.
In particular, TMA depends on the participation rates the different agents $i$ display, and therefore, it depends on the degrees $k_i$ or $q_i$ of the nodes.

\section{
\label{results}
Results
}

\begin{figure*}[ht]
\includegraphics[width=\textwidth]{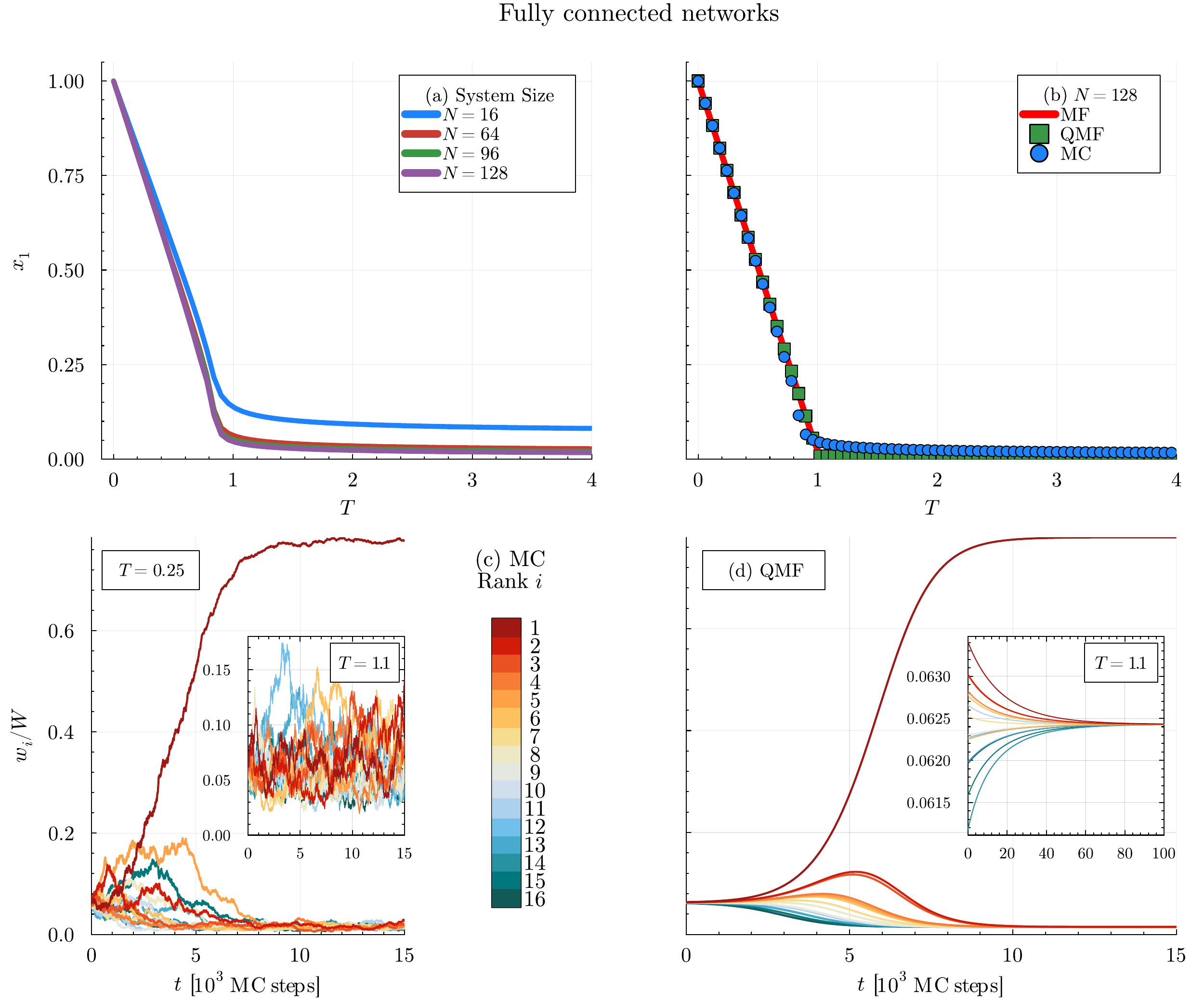}
\caption{
\label{fig:1}
Behavior of EYS model on fully connected networks.
{\bf a)} The expected relative wealth $x_1$ owned by the richest agent at the stationary regime is plotted vs temperature $T$ for different system sizes $N$, averaged over 64 MC simulations of the stochastic process of Eq.~\ref{eq:1} starting from uniform initial conditions.
{\bf b)} A comparison of $x_1$ vs $T$ for $N = 128$, showing results from MC simulations (blue circles), a single numerical simulation of the QMF approximation of Eq.~\ref{eq:8} starting from a uniform initial condition with added Gaussian noise (mean = 0, standard deviation = 0.01, green squares), and the MF prediction of Eq.~\ref{eq:12} (red solid line).
{\bf c)} In a system of $N=16$ agents, each line represents the time evolution of the $w_i/W$ for $i=1,...,16$ in a MC simulation  below the critical temperature, $T=0.25$.
The inset shows analogous results but above the critical temperature, $T=1.1$.
{\bf d)} Time evolution of the same system as in {\bf c)} but for a single numerical simulation of the QMF approximation at $T=0.25$ (main plot) and $T=1.1$ (inset).
}
\end{figure*}

\subsection{
\label{sec:fully_connected}
Fully connected networks
}

Let us begin with the case of fully connected networks to compare with previous results.
In this case, $k_i=q_i=N-1$, so $\kappa_{ij}=\avrg{c_{ij}}_{s|a}=2N^{-1}(N-1)^{-1}$ for all $j\neq i$ and $\kappa_{ii}=\avrg{c_{ii}}_{s|a}=0$ for all $i$ in both, IMA and IMB, and $\beta_i=\avrg{b_{i}}_{s|a}=\avrg{b_{j}}_{s|a}=\beta_j=2/N$ in TMA and TMB for two randomly selected agents.
Therefore, in the fully connected case, no distinction could be made between the different interaction and taxation modes.

As is already known~\cite{boghosian2017oligarchy} and confirmed in this work by numerical simulations of the stochastic process of Eq.~\ref{eq:1}, the system manifests a phase transition between two regimes.
For large values of $T$, the system exhibits a disordered phase where all agents have the same expected relative wealth $x_i=N^{-1}$.
For small values of $T$, a condensed phase emerges where one agent, say agent $i=1$, takes a macroscopic fraction of wealth, so $x_1=\bigo(1)$ and $x_i=\bigo(N^{-1})$ for all $i>1$.
Based on these observations, it makes sense to consider a MF approximation of the QMF approximation where $x_i=\bar{x}\leq x_1$ for all $i>1$, so $1=x_1+(N-1)\bar{x}$, so the ODE of Eq.~\ref{eq:8} simplifies to~\cite{klein2021meanfield}
\begin{eqnarray}
\label{eq:11}
\frac{N}{2}
\dot{x}_1
&=&
T N^{-2}
(1-N x_1)
+
(x_1-\bar{x})\bar{x}
\\
&=&
T N^{-2}
(1-N x_1)
+
\bigg(
x_1-\frac{1-x_1}{N-1}
\bigg)
\frac{1-x_1}{N-1}
\nonumber
\\
&=&
(1-N x_1)
\bigg(
\frac{T}{N^{2}}
-
\frac{1-x_1}{(N-1)^2}
\bigg)
.
\nonumber
\end{eqnarray}
From here and the condition $\dot{x}_1=0$, two equilibrium solutions are obtained:
$x_1^* = 1/N$ and 
\begin{equation}
\label{eq:12}
x_1^*=1-T(N-1)^2/N^2\approx 1-T
\end{equation}
as $N\to \infty$.

To study the stability of the equilibrium solutions, let $\epsilon := x_1-x_1^*$ and write
$$
\dot{\epsilon}
=
\dot{x}_1
=
f(x_1^*+\epsilon)
=
f(x_1^*)
+
f'(x_1^*)\epsilon
+
...
$$
where $f(x_1)$ is proportional to the r.h.s. of Eq.~\ref{eq:11}.
Since $f(x_1^*)=0$ because $x_1^*$ is an equilibrium solution, then the dynamics is stable if $f'(x_1^*)<0$, linearly marginal if $f'(x_1^*)=0$ and unstable if $f'(x_1^*)>0$.
The condition for the marginal case can be used to identify the critical point at which the condensation phase transition occurs. 
It predicts a critical temperature $T_c = N/(N-1) \approx 1$ as $N \to \infty$, with $x_1^*(T = T_c) = 1/N$ for both equilibrium solutions \( x_1^* \). 
However, the stability of these solutions differs: for $T < T_c$, $f'(x_1^*) > 0$, and for $T > T_c$, $f'(x_1^*) < 0$ when $x_1^* = 1/N$, while the reverse holds for the other equilibrium point. 
As expected, these results indicate that above $T_c$, all agents share approximately equally the same wealth $W/N$, while, in contrast, below $T_c$, a single oligarch emerges, owning a macroscopic fraction of the total wealth $W$.

Figure~\ref{fig:1}{\bf a)} shows results for MC simulations of the stochastic process of Eq.~\ref{eq:1} for different sizes $N$, which corresponds to the EYS model on fully connected networks.
Except when stated otherwise, the uniform state $x_i=1/N$ is used as the initial condition for all simulations.
The expected relative wealth of the richest agent $x_1$ in the steady state is plotted vs the temperature $T$.
The phase transition is confirmed and finite size effects are observed.
In Fig.~\ref{fig:1}{\bf b)} the MC simulations are compared against numerical predictions of the QMF approximation of Eq.~\ref{eq:8} and the analytical MF predictions derived from Eq.~\ref{eq:11}.
Here, a small Gaussian noise is added to the initial condition to break the symmetry.
As can be seen, the predictions agree up to finite size effects.
Examples of the microscopic dynamics from simulations of the MC and the QMF approximations of a system with $N = 16$ agents are shown in Figs.~\ref{fig:1}{\bf c)}~and~\ref{fig:1}{\bf d)}, respectively.
Moreover, the wealth of multiple agents increases with time during an initial transient period.
Note, however, this growth is short-lived.
Eventually, all agents with the exception of the richest begin to experience a decline in wealth.
This behavior arises because wealthier agents accumulate resources at the expense of their neighbors.
During the early stages, several secondary wealthy agents see their wealth grow while they extract resources from poorer neighbors.
Once these surrounding resources are exhausted, only the competition between the rich agents remain, ultimately favoring only the wealthiest.

\subsection{
\label{sec:MC_ER}
Erd\H{o}s-R\'enyi random networks
}

\begin{figure*}
\includegraphics[width=\textwidth]{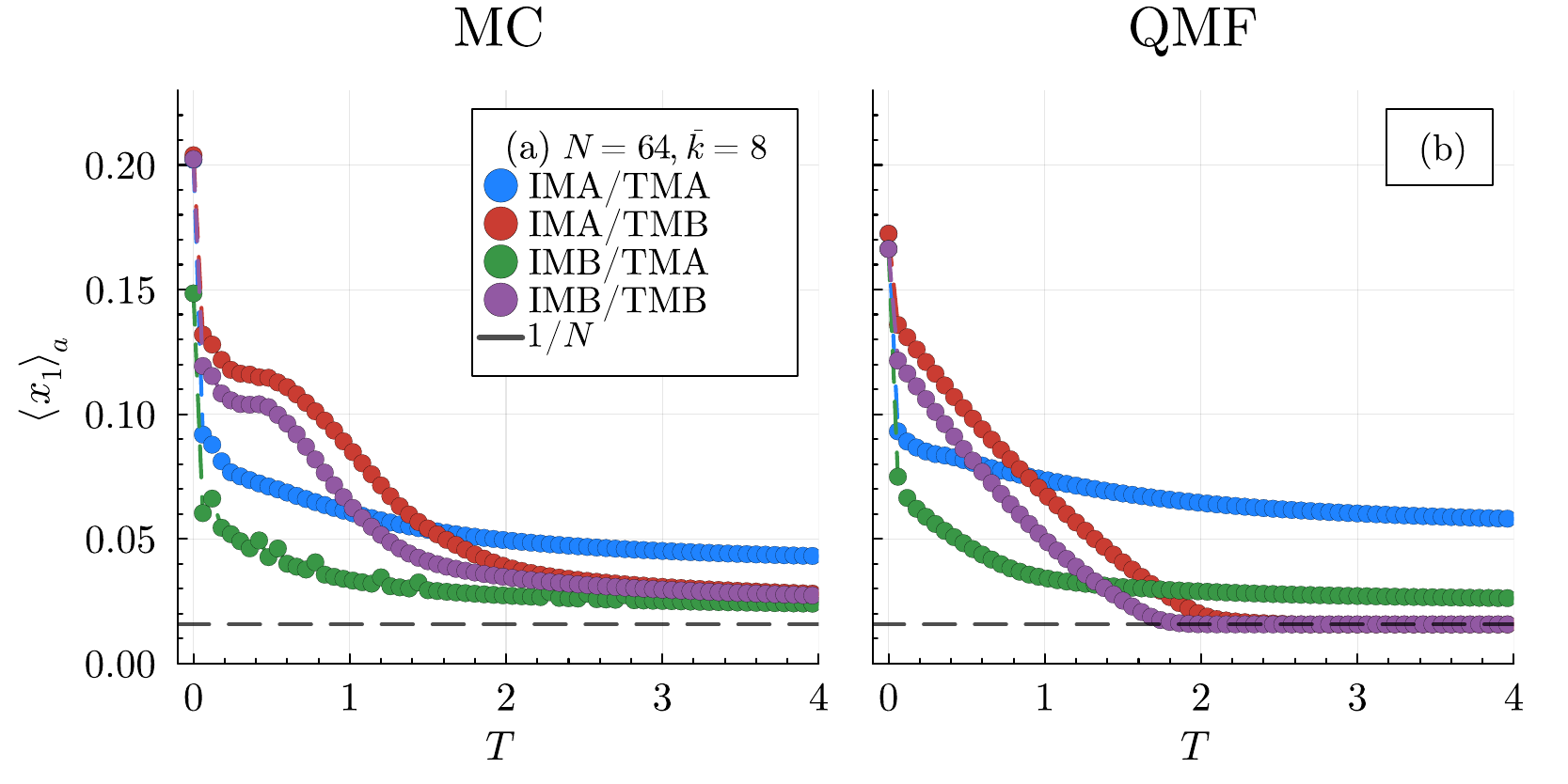}
\caption{
\label{fig:2}
The expected value $\avrg{x_1}_a$ of the relative wealth owned by the richest agent vs the temperature $T$, obtained at the stationary regime for different interaction and taxation modes, averaged over ER networks of $N=64$ nodes and average degree $\bar{k}=8$.
Panel {\bf a)} shows the average for 512 MC simulations and panel {\bf b)} over 2048 samples of QMF approximations.
}
\end{figure*}

Figure~\ref{fig:2}{\bf a)} compares the network-averaged relative wealth of the richest agent as a function of $T$ across all interaction and taxation modes, based on MC simulations of Eq.~\ref{eq:5} on Erd\H{o}s-R\'enyi random networks with $N = 64$ nodes and average degree $\bar{k} = 8$.
Figure~\ref{fig:2}{\bf b)} shows the corresponding results obtained using the QMF approximation of Eq.~\ref{eq:8}.
Important differences emerge when the interaction and taxation modes are varied.
Additionally, although the overall behavior is qualitatively similar, notable discrepancies also exists between the MC simulations and the QMF approximations---particularly in the TMB cases, where the MC curves display a plateau within the range $0<T<T_c$ that is not observed for QMF approximations.
Appendix~\ref{appA} sheeds light on the origin of this discrepancy, though a complete understanding of the phenomenon is still lacking.
The remainder of the analysis focuses on the IMA/TMA cases, where the QMF approximation exhibits the closest agreement with the MC results.

In the four studied cases of the EYS on ER networks, the largest values of $\avrg{x_1}_a$ are seen at $T=0$, where no redistribution is at work. 
Also, discontinuous decays to smaller values of $\avrg{x_1}_a$ are observed when transitioning from zero to small temperatures.
This phenomenon is exclusively a topological effect, as it is not observed in the case of fully connected networks.
At small but non-zero temperatures, the highest values of $\avrg{x_1}_a$ are observed for TMB, regardless of the interaction mode. Then, $\avrg{x_1}_a$ tends to the value $1/N$ as temperature increases, except for IMA/TMA where the richest agent maintains a significant advantage due to the network structure.
Interestingly, for ER networks, and at difference with the case of fully connected networks, the decay of $\avrg{x_1}_a$ vs $T$ in the condensed phase exhibits some differences between the MC simulations and the QMF approximations.

\begin{figure*}
\includegraphics[width=\textwidth]{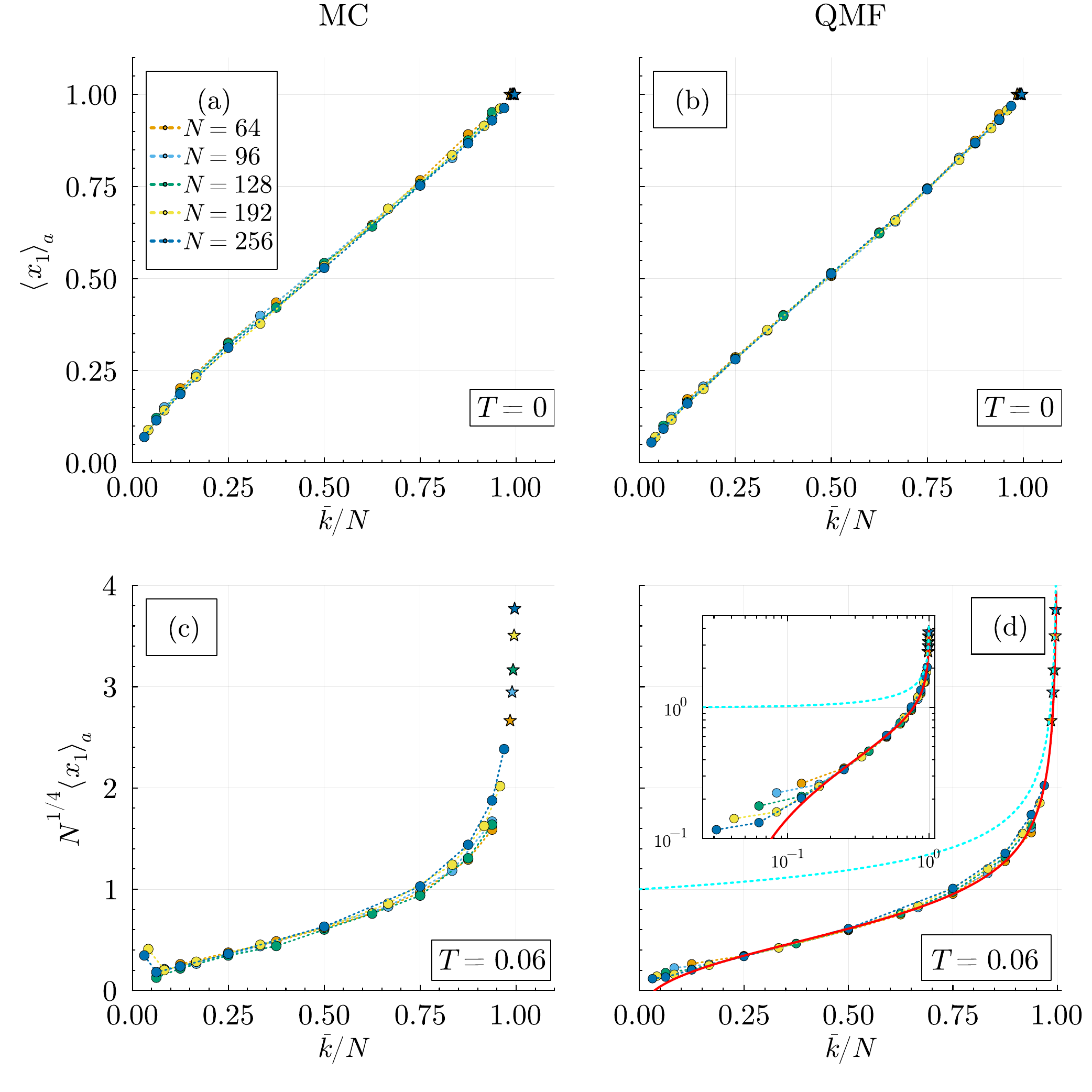}
\caption{
\label{fig:3}
Scaling of the network average of the expected relative wealth of the richest agent $x_1$ at the stationary regime, with respect to network sizes $N$ and average degrees $\bar{k}$, for MC simulations (left column) and QMF approximations (right column), at two specific temperatures, $T=0$ (top row) and $T=0.06$ (bottom row).
In all cases, the stars are obtained at $\bar{k}=N-1$, i.e. the fully connected case.
In panel {\bf d)}, the dashed cyan curve corresponds to the scaling $g_T(z)=(1-z)^{-1/4}$ for $z=(N-1)/N\lesssim 1$ and the red solid curve to the scaling $g_T(p)=-1/3+p^{1/3}(1-p)^{-1/4}$ for $p\approx  \bar{k}/N$.
In the inset, the same results are shown in log-log scale.
}
\end{figure*}

The numerical experiments show that, for all interaction and taxation modes, the value of $\avrg{x_1}_a$ grows with $\bar{k}$ but decays with $N$ at zero or small temperatures, $T\gtrsim 0$ (not shown).
In other words, there is a competition between $\bar{k}$ and $N$ for consolidating or weakening the condensed phase, respectively.
To better characterize this behavior, assume the scaling hypothesis 
\begin{equation}
\avrg{x_1}_a
(
\lambda^{\nu} \bar{k},
\lambda^{\mu} N
)
\approx 
\lambda
\avrg{x_1}_a(\bar{k},N)
\notag
\end{equation}
for arbitrary values of $\lambda$, $\bar{k}$, $N$ and zero or small temperatures.
Choosing the value of $\lambda$ such that $\lambda^{\mu} N=1$, the scaling relation
\begin{eqnarray}
\avrg{x_1}_a(\bar{k},N)
\notag
&\approx &
N^{1/\mu}
\avrg{x_1}_a(N^{-\nu/\mu}\bar{k},1)
\\
&=&
N^{1/\mu}
g_T(N^{-\nu/\mu}\bar{k})
\notag
\end{eqnarray}
is obtained, for some universal function $g_T$ that generally depends on $T$.
As shown in Fig.~\ref{fig:3}, a collapse of the curves $N^{-1/\mu}\avrg{x_1}_a(\bar{k},N)$ vs $N^{-\nu/\mu}\bar{k}$ is obtained for different system sizes $N$ while varying $\bar{k}$.
By visual inspection, it is found that the scaling occurs for $\nu=\mu=\infty$ such that $\nu/\mu=1$ when $T=0$ (Figs.~\ref{fig:3}~{\bf a)}~and~{\bf b)}) and for $\nu=\mu=-4$ when $T=0.06$ (Figs.~\ref{fig:3}~{\bf c)}~and~{\bf d)}).
In other words, the scaling
\begin{eqnarray}
\avrg{x_1}_a(\bar{k},N)
\approx
g_{T=0}(\bar{k}/N)
\approx
N^{-1}\bar{k}
\notag
\end{eqnarray}
is obtained at zero temperature, and the scaling
\begin{eqnarray}
\label{eq:13}
\avrg{x_1}_a(\bar{k},N)
\approx
N^{-1/4}g_{T\gtrsim 0}(\bar{k}/N)
\end{eqnarray}
is obtained at the small but non-zero temperature $T=0.06$.
These results hold for MC simulations and QMF approximations when uniform initial conditions are used.
An analogous scaling with the same exponents is found for the IMA/TMB case (not shown).
Observe that if $\avrg{x_1}_a = c$ for some constant $0 < c < 1$, then Eq.~\ref{eq:13} implies $\bar{k}/N \approx 1 - c^{-4}/N$.
Consequently, for non-zero temperatures, a non-zero average $\avrg{x_1}_a$ in the limit $N\to \infty$ is possible only for fully connected networks. 
This result aligns with the findings of B\"orgers and Greengard~\cite{borgers2024local}, who argued that global condensation occurs at $T=0$ exclusively for fully connected networks, while local condensation is observed otherwise.
Further details about the local wealth condensation observed in ER networks are discussed in Appendix~\ref{appB}.

For small but non-zero temperatures, we can characterize $g_T$ by considering the limiting case of a fully connected network, where the average degree $\bar{k}=N-1$. In this limit, the relation
$$
1 \approx x_1 \approx N^{-1/4}g_{T\gtrsim 0}\bigg(\frac{N-1}{N}\bigg)
$$
holds as $N\to \infty$ only if
$$
g_{T\gtrsim 0}\bigg(\frac{N-1}{N}\bigg) \approx N^{1/4}.
$$
Introducing the substitution $z=\frac{N-1}{N}$, this suggests the scaling
\begin{equation}
\label{eq:14}
g_{T\gtrsim 0}(z) \approx \bigg(\frac{1}{1-z}\bigg)^{1/4},
\end{equation}
which is visually confirmed by the dashed cyan line in Fig.~\ref{fig:3} for $z$ close to 1. 
Note, however, at smaller values of $z$, the prediction of Eq.~\ref{eq:14} fails.
A better fit is provided by the phenomenological proposition
$$
g_{T\gtrsim 0}(p) \approx -\frac{1}{3} + p^{1/3}(1-p)^{-1/4}
$$
across the entire range of $p\approx \bar{k}/N$, as confirmed by the solid red line of Fig.~\ref{fig:3} except at very small values of $p$ where the sparse network regime emerges.

\begin{figure*}[ht]
\includegraphics[width=\textwidth]{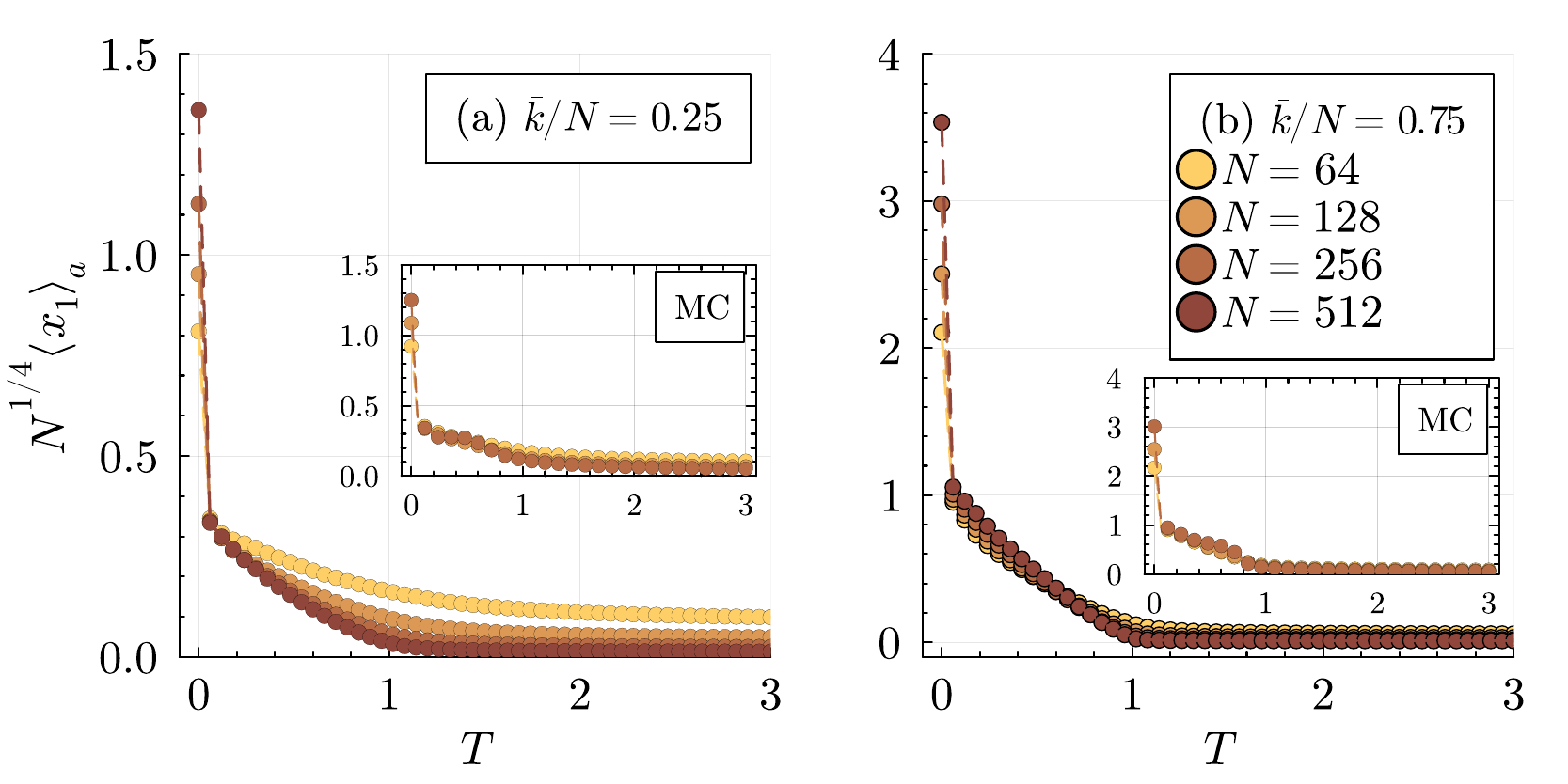}
\caption{
\label{fig:4}
Scaling of the network-averaged order parameter $N^{1/4}\langle x_1 \rangle_a$ as a function of temperature $T$, for systems of varying size $N$, grouped by link density. 
Panel {\bf a)} corresponds to $p \approx \bar{k}/N = 0.25$, and panel {\bf b)} to $p \approx 0.75$. 
The main panels display results obtained using QMF approximations, while the insets show the corresponding results from MC simulations (up to $N = 256$).
}
\end{figure*}

To understand how previous scaling behavior is affected by $T$ beyond the regimen of small temperatures, $N^{1/4}\avrg{x_1}_a$ is plotted vs $T$ in Fig.~\ref{fig:4}.
At relatively small values of the link density $p$, finite size effects are observed in Fig.~\ref{fig:4}{\bf a)}.
As $p$ increases, these finite size effects diminishes as shown Fig.~\ref{fig:4}{\bf b)}.
These observations are common to QMF approximations (main panels) and MC simulations (insets).
Additionally, the critical point $T_c=1$ can be appreciated as $N\to \infty$.

A key observation in the stationary regime is the lack of connections among rich agents. 
This can be intuitively understood as a consequence of competition: connected wealthy agents engage in interactions that ultimately lead to the financial ruin of one of them. 
To analyze this phenomenon more formally, a MF approximation is developed for the EYS model on ER random networks.
Namely, assume that $k_i=q_i\approx \bar{k}=\bar{q}$ for all $i$, so $\avrg{\beta_i}_a=\avrg{b_{i}}_{sa}\approx 2/N$ for TMA and TMB, and $\avrg{\kappa_{ij}}_a=\avrg{c_{ij}}_{sa} \approx 2a_{ij}/(\bar{k}N)$ for IMA and IMB, so Eq.~\ref{eq:8}  takes the form
\begin{equation}
\label{eq:15}
\frac{N}{2}
\dot{x}_i
\approx
\frac{T}{N^2}
(1-Nx_i)
+
\sum_j
\frac{a_{ij}}{\bar{k}}
(x_i-x_j)
\min\{x_i,x_j\}
.	
\end{equation}
Next, assume that the population is divided between $R$ rich agents with wealth $u\approx \avrg{x_1}_a\approx ...\approx \avrg{x_R}_a$ and $N-R$ poor agents with wealth $v\approx \avrg{x_{R+1}}_a\approx ...\approx \avrg{x_N}_a$, where $u>1/N>v$.
In this way, the total normalized wealth is $1=Ru+(N-R)v$ and, from Eq.~\ref{eq:15}, the following ordinary differential equation for the expected wealth of rich agents is obtained
\begin{eqnarray}
\label{eq:16}
\frac{N}{2}
\dot{u}
&\approx &
\frac{T}{N^2}(1-Nu)
+
(u-v)v
\\
&=&
(1-Nu)
\bigg(
\frac{T}{N^2}
-
\frac{1-Ru}{(N-R)^2}
\bigg)
\nonumber
\end{eqnarray}
where the $k_i\approx \bar{k}$ terms in the summation cancel out with the factor $1/\bar{k}$.
It is important to remark that $R$ is actually an unknown function of $T$, $N$, $\bar{k}$ and $u$.
For a given $T$, the equilibrium condition $\dot{u}=0$ derived from Eq.~\ref{eq:16} reveals two solutions: $u^*=1/N$ and 
\begin{equation}
\label{eq:17}
u^* 
=
\frac{1}{R^*}
\bigg(
1
-
T
\frac{(N-R^*)^2}{N^2}
\bigg)
\end{equation}
where $R^*=R(T,N,\bar{k},u^*)$.
Note that for $R^*=1$, Eqs.~\ref{eq:16}~and~\ref{eq:17} simplify to those of the MF theory over fully connected networks.
The condition of marginal linear stability $\dot{u}=0$ implies the critical temperature 
\begin{equation}
T_c=N/(N-R^*).
\notag
\end{equation}
for both values of $u^*=u(T,N,\bar{k},R^*)$.
Additionally, $u^*=1/N$ for $T=T_c$.
For $T>T_c$ the fix point $u^*=1/N$ is stable and the fix point of Eq.~\ref{eq:17} is unstable, while the opposite holds for $T<T_c$. 
In other words, a condensation phase transition occurs at $T_c$.
The stable equilibrium solution is $u^*=1/N$ above $T_c$ and is that of Eq.~\ref{eq:17} below $T_c$.

\begin{figure*}[ht]
\includegraphics[width=\textwidth]{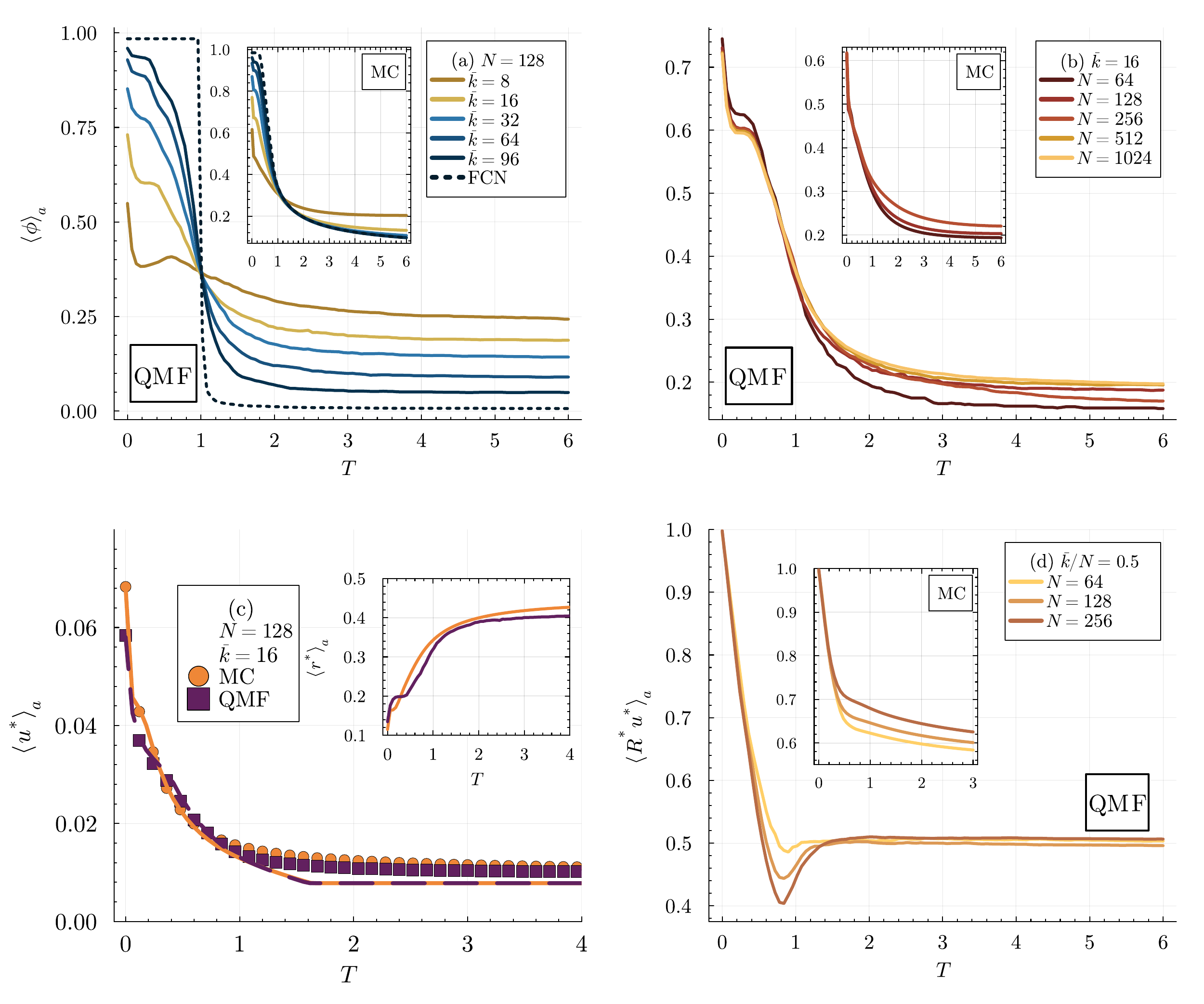}
\caption{
\label{fig:5}
{\bf a)} The steady state network average of the order parameter $\phi$ defined in Eq.~\ref{eq:19} is plotted as a function of temperature $T$ for networks of $N=128$ nodes and different average degrees $\bar{k}$, using IMA and TMA, for QMF approximations. 
In the inset, analogous results for for MC simulations.
{\bf b)} Similar to panel {\bf a)} and corresponding inset, but fixing the average degree $\bar{k}=8$ while varying the network size $N$.
{\bf c)} 
The network average of the average wealth $u^*$ of the rich agents is plotted as a function of $T$ by direct calculation with MC simulations (orange circles) and QMF approximations (purple symbols) for $N=128$ and $\bar{k}=8$.
For comparison, corresponding predictions from Eq.~\ref{eq:17} obtained by direct calculations of network averages of the number $R^*$ of rich agents are included for MC simulations and QMF approximations (solid lines of respective colors).
{\bf d)} The network average of the aggregated wealth of the rich agents $R^*u^*$ is plotted as a function of the temperature $T$ for different values of $N$, keeping fixed the link density $\bar{k}/N=0.5$, for QMF approximations (main plot) and MC simulations (inset).
}
\end{figure*}

Building on the previous considerations, let $R_{\max}$ denote the maximum number of non-adjacent (i.e., mutually non-connected) rich agents that a network can sustain. 
Determining $R_{\max}$ is equivalent to finding a maximal independent set in the network~\cite{dallasta2009statistical}, and its value depends strongly on the network topology. 
For example, in a one-dimensional ring, poor and rich agents can alternate, yielding $R_{\max} \approx N/2$. 
In contrast, for a fully connected network, the only possible independent set consists of a single node, so $R_{\max} = 1$.
In any case, the expected fraction of rich agents, $r := R/N$, is bounded by $r \leq r_{\max} := R_{\max}/N$.
In the high-temperature limit $T \to \infty$, this fraction approaches $r \to 1/2$, while in the dense network limit $\bar{k} \to N - 1$, we have $r \to 0$.
For Erdős–Rényi networks, $R_{\max}$ is expected to decrease with increasing link density $p \approx \bar{k}/N$. In particular, for $N \gg R \gg 1$ and sufficiently large $p$, it satisfies the asymptotic relation~\cite[Theorem 7.1]{janson2000random}:
\begin{equation}
\label{eq:18}
R_{\max} \approx \frac{2 \ln N}{\ln \left( 1/(1 - p) \right)}.
\end{equation}
Based on this, it is natural to introduce the order parameter
\begin{equation}
\label{eq:19}
\phi := 1 - 2r,
\end{equation}
which captures the imbalance between poor and rich agents. By construction, $\phi \to 0$ as $T \to \infty$, and $\phi \approx 1$ in the low-temperature and high-connectivity limit $T \to 0$, $\bar{k} \to N - 1$.
These predictions for $\phi$ are corroborated by both MC simulations and QMF approximations, as shown in Figs.~\ref{fig:5}~{\bf a)}~and~{\bf b)}.
In particular, Fig.~\ref{fig:5}{\bf a)} reveals a crossover at a characteristic temperature $T = T_c$, which appears consistently across different values of $\bar{k}$.
Moreover, while the transition is continuous in MC simulations, it becomes discontinuous in the QMF approximation as $\bar{k} \to N - 1$, likely due to the mean-field treatment.
The behavior of the time-averaged order parameter $\langle \phi \rangle_a$ also reflects the interplay between connectivity $\bar{k}$ and system size $N$.
For $T < T_c$, $\langle \phi \rangle_a$ increases with $\bar{k}$ and decreases slightly with $N$; the opposite trend is observed for $T > T_c$.
These trends are consistent with the approximation $r_{\max} \approx (2/\bar{k}) \ln N$ derived from Eq.~\ref{eq:18} for $\bar{k} \ll N$ and $T \approx 0$, which predicts a logarithmic growth with $N$ and an inverse dependence on $\bar{k}$.

Figure~\ref{fig:5}{\bf c)} shows the network average of $u^*$ as a function of temperature $T$, based on MC simulations (orange circles) and QMF approximations (purple squares). 
For comparison, the predictions of Eq.~\ref{eq:17}, evaluated using the network-averaged number $R^*$ of rich agents obtained from simulations, are also plotted (colored lines matching the corresponding data).
The inset displays the network average of the fraction of rich agents, $r^* = R^*/N$, as a function of $T$, from which the values of $R^*$ are derived.
As temperature increases, $u^*$ decreases up to a crossover point around $T \approx T_c$, beyond which it remains approximately constant.
This behavior is mirrored by $\langle r^* \rangle_a$, which starts near $r_{\max} = R_{\max}/N$ for $T \gtrsim 0$, rises sharply up to $T_c$, and then increases more gradually toward the asymptotic value $1/2$.
Overall, the various curves show good quantitative agreement, supporting the consistency between simulation results, QMF approximations, and theoretical predictions.

In Fig.~\ref{fig:5}{\bf d)} network averages of the cumulative wealth of the rich agents, $u^*R^*$ are plotted as a function of $T$ for different network sizes $N$ and a fixed link density $p\approx \bar{k}/N$ for QMF approximations (main plot) and MC simulations (inset).
A clear change in behavior is observed just below $T_c$, in agreement with the existence of a phase transition.
In particular, for QMF approximations, there is a non-monotonous behavior where the minimum of $u^*R^*$ drops with $N$ because $R^*$ decays faster than the growth of $u^*$ at such point.
This behavior is not observed in MC where $R^*$ tends to overestimate the effective number of rich agents due to stochastic fluctuations.

\begin{figure*}[ht]
\includegraphics[width=\textwidth]{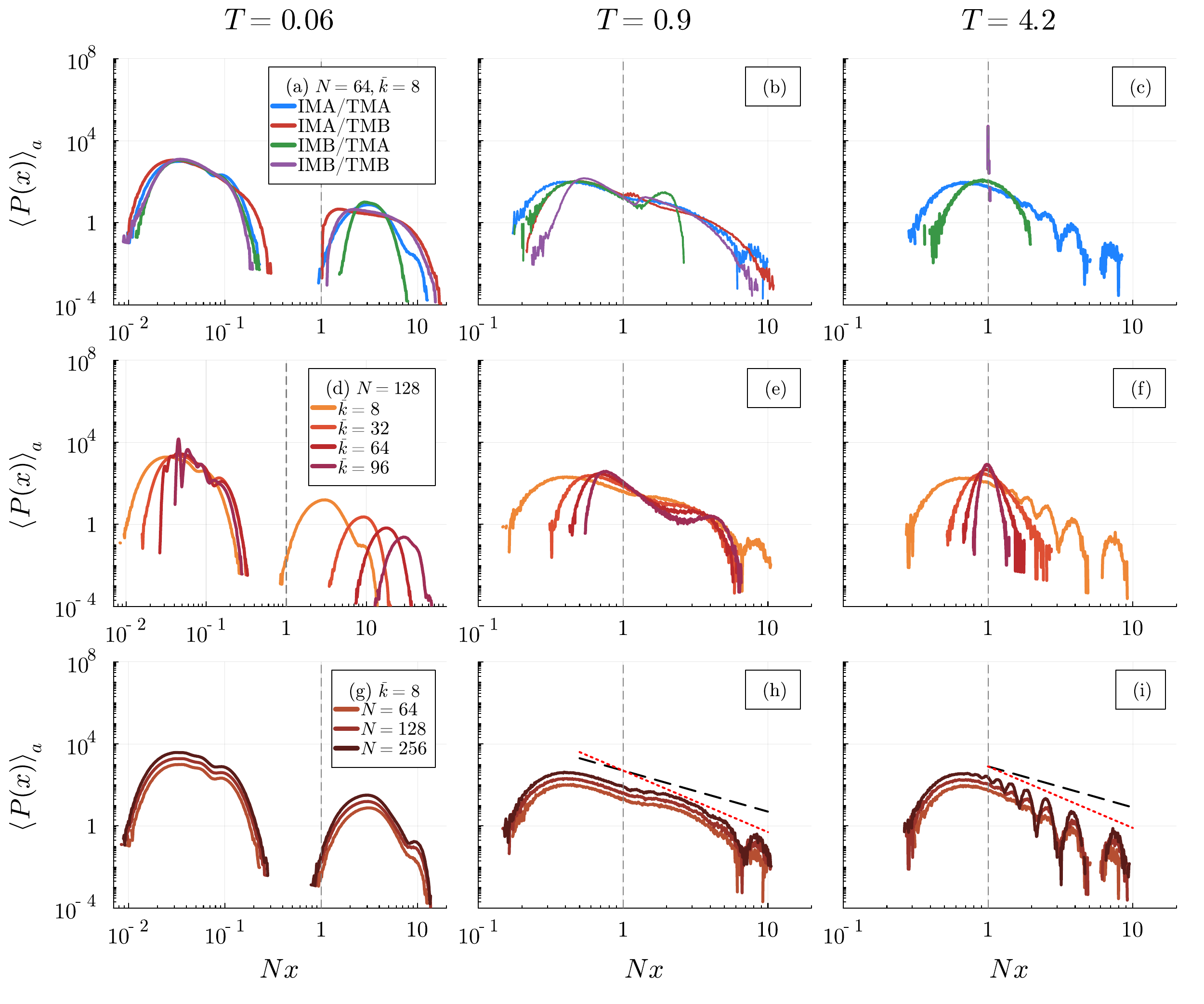}
\caption{
\label{fig:6}
Network average of the normalized wealth distribution $P(x)$ in the equilibrium state, obtained from QMF approximations for system at different combinations of $N,\bar{k}$ and $T$, and different interaction and taxation modes.
The left column shows the results for a temperature near zero, the middle column for an intermediate temperature near $T_c$, and the right column for a high temperature above $T_c$.
\textbf{a)-c)} Wealth distribution for a system with $N=64$ and $\bar{k}=8$, comparing different interaction/taxation modes.
\textbf{d)-f)} Wealth distribution for a system with $N=128$ and IMA/TMA, for different values of the average degree $\bar{k}$.
\textbf{g)-i)} Wealth distribution for a system with $\bar{k}=8$ and IMA/TMA, showing size effects.
For visual guidance, the black dashed and red dotted lines indicate power-law decays with exponents $-2$ and $-3$, respectively.
}
\end{figure*}

In Fig.~\ref{fig:6} the network average of the distribution of normalized wealths $P(x)$ is shown for QMF approximations in different scenarios.
The top row shows distributions for all interaction and transaction modes for relatively sparse networks.
In the middle and bottom row, distributions for the IMA/TMA modes are shown for varying degrees $\bar{k}$ and network sizes $N$, respectively.
In the left column, the distribution is shown for the condensed phase. 
In all cases two modes or peaks are observed, one for poor agents and another for rich agents (Figs.~\ref{fig:6}~{\bf a)},~{\bf d)}~and~{\bf g)}).
This result has also been observed in the MC simulations. 
In the middle column, distributions obtained for temperatures near or below the critical point are shown.
Broad distribution are observed, except for the IMB/TMA case, where there is a sharp cutoff at the tail of the curve (Fig.~\ref{fig:6}{\bf b)}).
The distribution broadens as the average degree $\bar{k}$ decreases for IMA/TMA (Fig.~\ref{fig:6}{\bf e)}).
Interestingly, this effect is also observed at high temperatures (Fig.~\ref{fig:6}{\bf f)}).
However, as seen in Fig.~\ref{fig:6}{\bf c)}, this is not the case for the other modes, especially for TMB where the distribution is highly concentrated around $x=1/N$.
Finally, as shown in Figs.~\ref{fig:6}~{\bf g)},~{\bf h)}~and~{\bf i)}, the distributions suffer little variation with network size.

\section{
\label{conclusion}
Conclusion
}

This work introduces a networked variant of the Extended Yard Sale model proposed by Boghosian et al.~\cite{boghosian2017oligarchy}.
The model is analyzed on Erd\H{o}s-R\'enyi random networks, combining Monte Carlo simulations with two theoretical approximations: a Quenched Mean Field approximation, which significantly reduces computational complexity, and a Mean Field theory that emphasizes the role of the number of rich agents in the system’s dynamics.

The model supports various interaction and taxation schemes, some of which display markedly different behaviors.
Despite these differences, a scaling analysis reveals that none of the variants exhibit global wealth condensation.
Instead, all variants show a form of local wealth condensation, as described by B\"orgers and Greengard~\cite{borgers2024local}, even in the presence of wealth redistribution mechanisms. Furthermore, consistent with the original EYS model on fully connected networks, all variants undergo a continuous phase transition when the competing effects of wealth redistribution and Wealth Attained Advantage reach a balance.

This work advances the broader understanding of Random Asset Exchange models. 
Under the hypothesis that economic systems may operate optimally near the critical point of a continuous phase transition, the results may inform the development of more effective redistribution policies.
Future research could extend the analysis to more realistic network topologies--such as scale-free, small-world, or higher-order networks~\cite{perotti2025analysis}--to test the robustness and universality of the observed phenomena.

\begin{acknowledgments}
The authors acknowledge partial support from CONICET under grant PIP2021-2023 No. 11220200101100 and SeCyT, Universidad Nacional de C\'ordoba (UNC), Argentina, helpful discussions to O.V. Billoni, F. Laguna, L. Giordano, S. Bustingorry, J. Almeira, S. A. Cannas and R. Iglesias, and computational resources from UNC Supercómputo (CCAD-UNC), which are part of SNCAD, Argentina.
\end{acknowledgments}

\appendix

\section{
Nonequilibrium behavior
\label{appA}
}

This appendix provides insights into the origin of the discrepancies between the MC simulations and the QMF approximations of the EYS model on ER networks at low temperatures, where the TMB cases display a plateau in Fig.~\ref{fig:2}.

\begin{figure}
\begin{center}
\includegraphics[width=\linewidth]{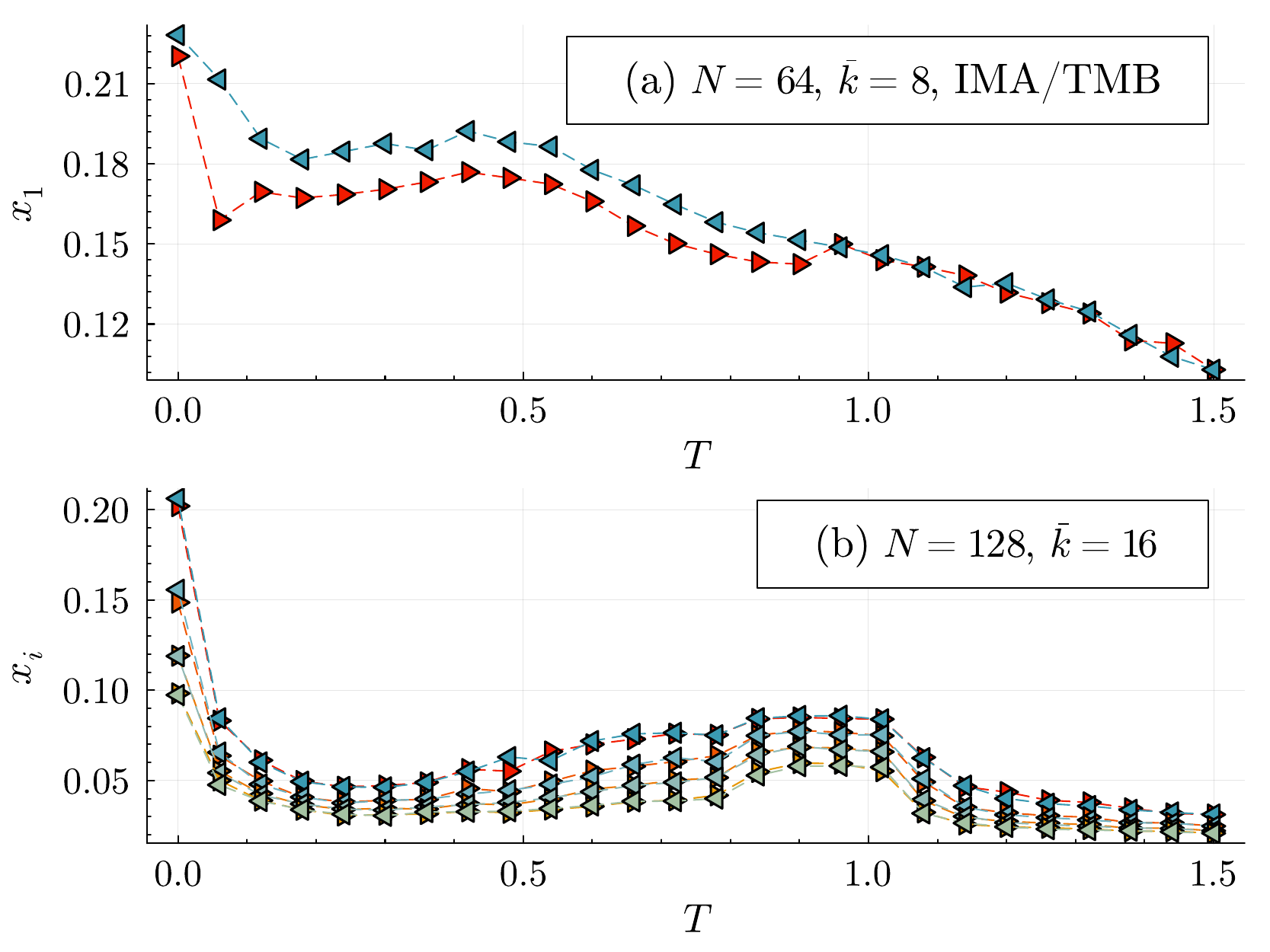}
\end{center}
\caption{
\label{fig:A1}
Hysteresis analysis of the EYS model on ER networks with average connectivity $\bar{k}/N=0.25$, for the IMA/TMB variant.
All data represent averages over 64 MC realizations on a single network.
{\bf a)} Hysteresis loop for $x_1$ (relative wealth of the richest agent) as a function of temperature $T$.
Red left-pointing triangles indicate the heating phase (starting from $T=0$ and uniform initial conditions),
while blue right-pointing triangles correspond to the cooling phase.
Dashed lines serve as visual guides.
{\bf b)} Average relative wealth $x_i$ of top-ranked agents as a function of $T$.
From top to bottom, each pair of curves corresponds to first, second and subsequent richest agents.
Coloring and symbols follow the same convention as in panel {\bf a)}.
}
\end{figure}

\begin{figure}
\begin{center}
\includegraphics[width=\linewidth]{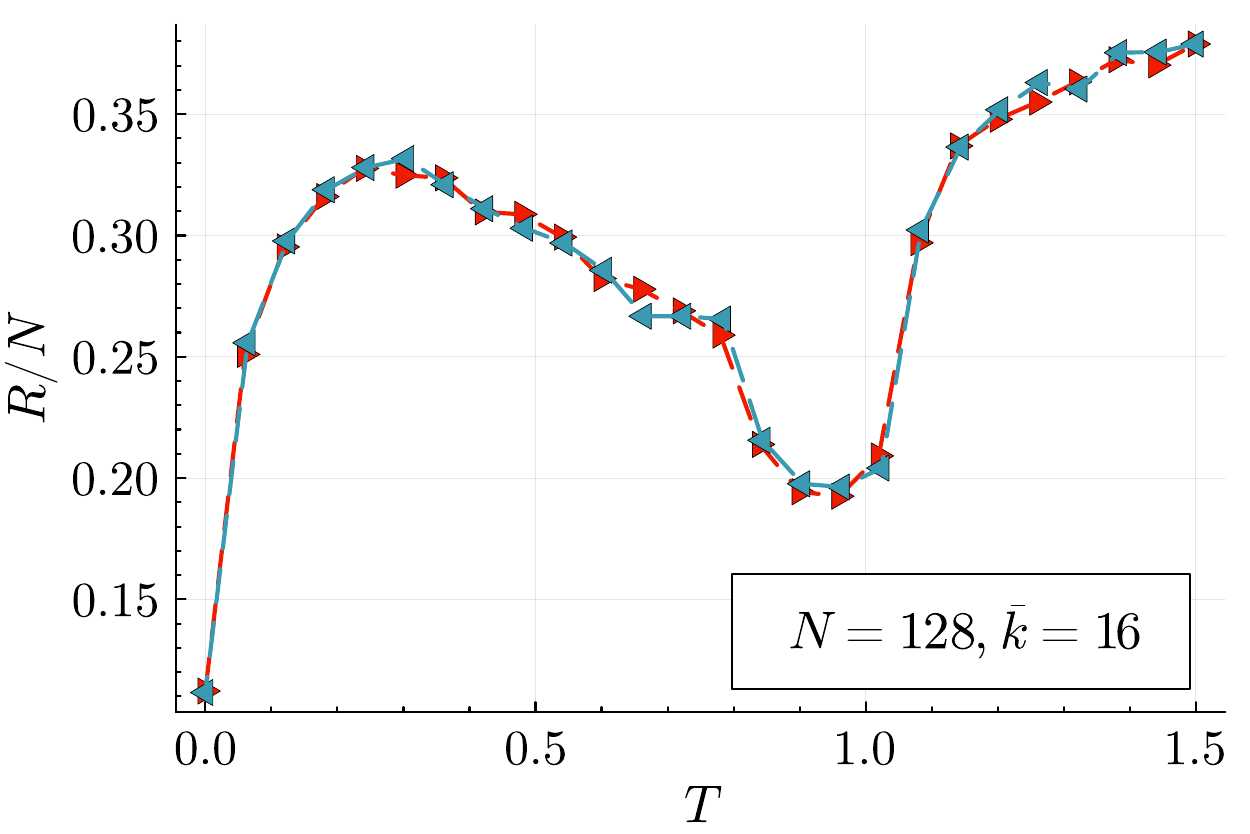}
\end{center}
\caption{
\label{fig:A2}
The fraction of rich agents $R/N$ is plotted as a function of temperature $T$ in the hysteresis analysis of the IMA/TMB case.
Each curve represents an average over 64 MC realizations on a single ER network.
As in Fig. \ref{fig:A1}{\bf a)}, red left-pointing triangles indicate the heating phase, while blue right-pointing triangles correspond to the cooling phase.
Recall that an agent $i$ is considered rich if its wealth exceeds the average wealth, i.e. $x_i>1/N$.
}
\end{figure}

In Fig.~\ref{fig:A1}{\bf a)} an hysteresis cycle is shown for MC simulations of the IMB/TMB case for one ER network of size $N=64$ and average degree $\bar{k}=8$.
The cycle begins at $T=0$ starting from an uniform initial condition, $x_i=1/N$.
As can be seen in the figure, a loop  between the temperatures $T=0$ and $T_c$, where the curve during the heathing phase (right triangles) goes mostly below the curve of the cooling phase (left triangles).
This results differs from the observed in Fig.~\ref{fig:2}, indicating that the system dynamics is not at thermodynamic equilibrium and, therefore, the initial condition plays an important role.
The area within the loop varies significantly with the network sample, but decreases with $N$ possibly due to the self-averaging of the network heterogeneities (compare Figs.~\ref{fig:A1}~{\bf a)}~and~{\bf b)}).
Interestingly, the curves exhibit a region where the average share of wealth $\avrg{x_1}_a$ of the richest agent grows with $T$.
This counterintuitive behavior is shared by the other rich agents (see Fig.\ref{fig:A1}{\bf b)}).
Moreover, as shown in Fig.~\ref{fig:A2}, the opposite tendency is observed for the number of rich agents.
These results, together with the absence of the plateau in QMF approximations, suggest that local wealth condensation is a necessary but not sufficient condition for the discrepancy to occur.

\begin{figure}
\begin{center}
\includegraphics[width=\linewidth]{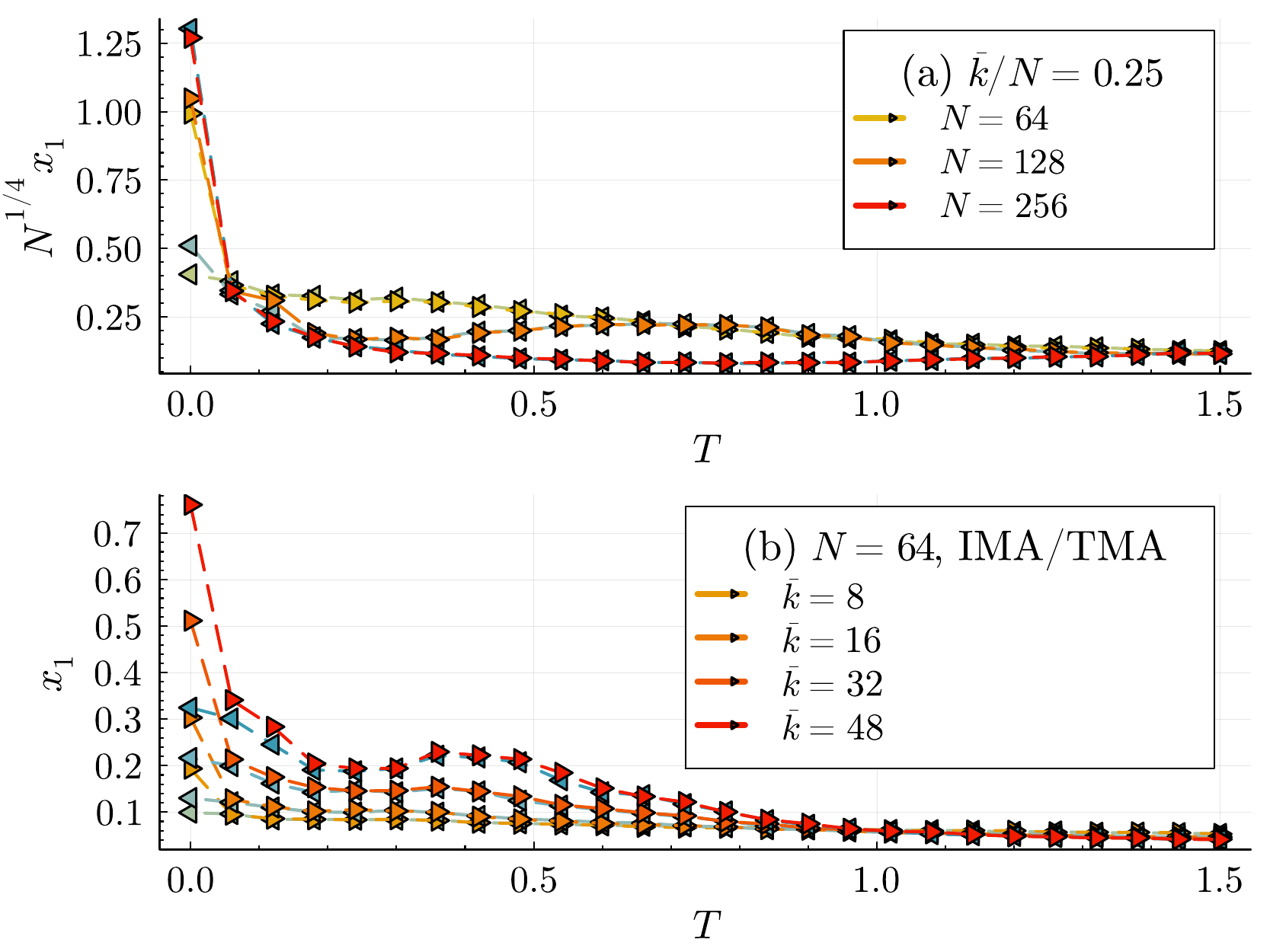}
\end{center}
\caption{
\label{fig:A3}
Scaling of hysteresis effects with connectivity and system size.
{\bf a)} Hysteresis cycles of $N^{1/4}x_1$ for different system sizes $N$, with fixed mean connectivity at $\bar{k}/N \approx 0.25$.
{\bf b)} Hysteresis cycles of $x_1$ for $N=64$ and varying average degree $\bar{k}$.
Data are taken from the same simulations as in Fig.~\ref{fig:A2}{\bf b)}.
}
\end{figure}

As depicted in Fig.~\ref{fig:A3}, a similar behavior is also found for IMA/TMA, but with a significantly smaller hysteresis loop.
More specifically, the area within the loop decreases with $N$ like in the TMB cases (see Fig.~\ref{fig:A3}{\bf a)}), but grows with the average degree $\bar{k}$ (see Fig.~\ref{fig:A3}{\bf b)}) where, to facilitate the visual comparison for the different sizes $N$, the curves are adjusted by the scaling of Eq.~\ref{eq:13}.

Previous results do not affect the scaling found at $T=0$ and $T\gtrsim 0$ in the experiments based on the uniform initial condition, $x_i=1/N$.

\section{
Local wealth condensation
\label{appB}
}

\begin{figure}
\includegraphics[width=\linewidth]{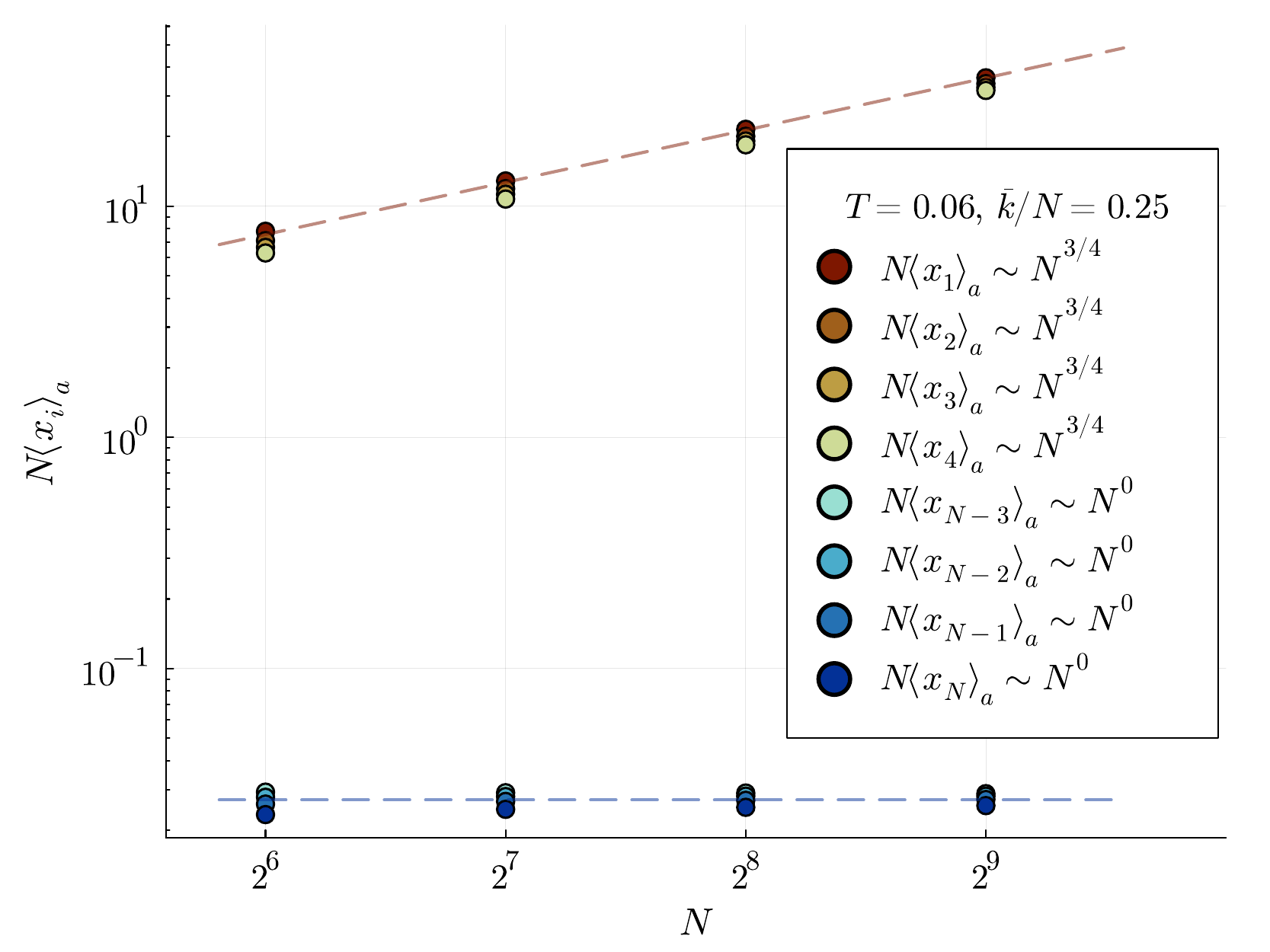}
\caption{
\label{fig:B1}
Scaling of the average wealth $\avrg{w_i}_a=N\avrg{x_i}_a$ with the system size $N$ for the top four richest agents and the four bottom poorest agents, for ER networks with average connectivity given by $\bar{k}/N=0.25$ and the IMA/TMA case.
Each circle is obtained averaging over 32 ER networks and 64 QMF approximations per network.
The dashed lines are guides to the eye following the scaling $\sim N^{3/4}$ for the richest agents (in red) and $\sim 1$ for the poorest (in blue).
}
\end{figure}

In the case of global wealth condensation, as seen in fully connected networks, the richest agent accumulates a macroscopic fraction of the total wealth, i.e., $\langle w_1 \rangle_a \sim N$, while the remaining agents each hold only a microscopic amount, $\langle w_i \rangle_a \sim 1$ for $i > 1$.
In contrast, local wealth condensation in ER networks prevents any single agent from accumulating a macroscopic share of the total wealth. Instead, a clear separation emerges between a population of rich agents and one of poor agents.
As shown in Fig.~\ref{fig:B1}, the average wealth of the rich agents increases with system size as $\langle w_{\mathrm{rich}} \rangle_a \sim N^{3/4}$, while the wealth of the poor agents remains constant, $\langle w_{\mathrm{poor}} \rangle_a \sim 1$.

%

\end{document}